\documentclass[prb,eqsecnum,preprint,showpacs]{revtex4}
\usepackage{amssymb,amsmath,bm}
\font\bba=msbm10 scaled 1080
\font\bbb=msbm8 
\font\bbc=msbm6 
\newfam\bbfam
\textfont\bbfam=\bba
\scriptfont\bbfam=\bbb
\scriptscriptfont\bbfam=\bbc
\def\bb{\fam\bbfam\bba}
\def\R{{\bb R}}

\begin{document}
\title{Random Walks on  Hyperspheres of Arbitrary Dimensions}
\author{Jean-Michel Caillol}
\affiliation{Laboratoire de Physique Th\'eorique \\
UMR 8267, B\^at. 210 \\
Universit\'e de Paris-Sud \\
91405 Orsay Cedex, France}
\email{Jean-Michel.Caillol@th.u-psud.fr}
\date{\today}
\begin{abstract}
We consider random walks on the surface of the sphere $S_{n-1}$
($n \geq 2$) of the
$n$-dimensional Euclidean space $E_n$, in short a hypersphere.
By solving the diffusion equation in $S_{n-1}$ we show that the usual law 
$\langle r^2 \rangle \varpropto t $ valid in $E_{n-1}$ should be replaced
in $S_{n-1}$ by the generic law
$\langle \cos  \theta \rangle \varpropto \exp(-t/\tau)$,
where $\theta$
denotes the angular displacement of the walker. More generally one has 
$\langle C^{n/2-1}_{L}\cos( \theta  )\rangle \varpropto \exp(-t/ \tau(L,n))$ where
$C^{n/2-1}_{L}$  a Gegenbauer polynomial. Conjectures concerning  random
walks on a fractal inscribed in $S_{n-1}$ are given tentatively. 
\end{abstract}
\pacs{05.40.Fb,05.40.Jc}
\maketitle
\newpage
\section{Introduction}
\label{intro}
We consider a random walk (RW)
on the surface of a sphere $S_{n-1}$ of the real
Euclidean space $E_n$ of dimension $n$, a hypersphere for short.
Let $O$ be the center of the hypersphere
and $R$ its radius, $S_{n-1}$ is thus defined as the set of points $M
\equiv(x_1,\ldots,x_n)$  of $E_n$ 
such that $\sum_{i=1}^{n} x_{i}^2=R^2$.
If the legs of the walker are much smaller than the radius $R$ of the sphere
then the diffusion process can be  described by a 
continuous diffusion equation, i.e.
\begin{equation}
\label{diff}
\mathcal{D} \rho(M,t) \equiv\left(\frac{\partial}{\partial t} 
-D\Delta^{{S}_{n-1}} \right)
 \rho(M,t)=0 \; ,
\end{equation}
where $\rho(M,t)$ is the density probabibily for the walker to be at point $M$
at time $t$. $D (>0)$ is the diffusion coefficient and
$\Delta^{S_{n-1}}$ denotes the Laplace-Beltrami operator in space $S_{n-1}$.
Explicit initial conditions will be specified later.
Note that for imaginary times eq.\ (\ref{diff}) identifies with
the Schr\"odinger equation of
a free particle of $S_{n-1}$, alternatively this equation can be considered as 
describing the diffusive rotational motion of a $n$-dimensional linear polar
molecule.

The solution of eq.\ (\ref{diff}) in the  case $n=3$ is known since the
work of Debye on Brownian rotors. \cite{Debye,Berne}
In this paper we extend Debye's proof to arbitrary
dimensions $n$ and show that a salient feature of a RW on a hypersphere is the
validity  of the generic law
 \begin{equation}
 \label{cos}
\langle \bm{\xi}(0) \cdot \bm{\xi}(t) \rangle \equiv 
\langle \cos \theta\left( t \right)  \rangle =
\exp \left( -D (n-1)t/R^2  \right) \; ,
\end{equation}
where  the unit vector $\bm{\xi}=\bm{OM}/R$ denotes the orientation of vector
$\bm{OM}$ in $E_n$. With the north pole of  $S_{n-1}$ chosen to be 
$M(t=0)$ the angle $\theta$ in eq.\ (\ref{cos}) 
is thus the colatitude of point $M$ in spherical coordinates. The
brackets $\langle \ldots \rangle$   denote
a spatial average with 
weight $\rho(M,t)$, and the dot in the LHS 
is the usual  scalar product in $E_n$.  
The result\ (\ref{cos}), given without proof in ref.~\onlinecite{Jullien},
is not surprising; indeed as noted by Jund \textit{et al.}
when $Dt/R^2 \ll 1$ one can expand
both sides of eq.\ (\ref{cos}) yielding for the projection $\bm{x}$ of vector
$\bm{OM}$ in the euclidean plane $E_{n-1}$ tangent to $S_{n-1}$ at point
$M_0$ to the 
behavior 
$\langle \bm{x}^2\rangle \sim R^2 \langle \theta^2\rangle \sim 2t(n-1)D $
characteristic 
of a random walk in an euclidean space of $(n-1)$ dimensions.

Recently Nissfolk \textit{et al.} have given a solution of\ (\ref{diff}) in the
special case $n=4$.\cite{Nissfolk}
Their expression for $\rho(M,t)$ is different from that which 
we derive in sec.\
\ref{general} and, in particular,
does not allow to recover easily eq.\ (\ref{cos}). The equivalence
of the two solutions is  established in the last section of this paper.

Our paper is organized as follows. In next section\ \ref{simple} we give an
elementary proof of eq.\ (\ref{cos}). The expression of the Green's function of
eq.\ (\ref{diff})  is derived in section\ \ref{general} from which 
eq.\ (\ref{cos}) can deduced. The
solution, inspired from the seminal paper of Debye on 3D rotors 
\cite{Debye,Berne},
requires the whole machinery of hyperspherical
harmonics and the use of some properties of the rotation group
$SO(n)$ \cite{Vilenkin}.  Finally in
section\ \ref{S3} we consider in more details the case $n=4$. Our
solution of\ (\ref{diff}) for $n=4$  and that proposed by 
Nissfolk \textit{et al.} \cite{Nissfolk} are shown to be identical. 
A preliminary account of the present work, however devoted to the special
case $n=4$, can be found in ref.~\onlinecite{Caillol1}.

\section{An elementary proof of 
 $\langle \cos  \theta \rangle \varpropto \exp(-t/\tau)$}
\label{simple}
We suppose that the walker it standing at some point
 $\bm{OM}_0=R \bm{\xi}_0$ of $S_{n-1}$
at time $t=0$. At subsequent times its dynamics is governed by 
eq.\ (\ref{diff}) and we study the time behavior of the random variable
$\cos \theta \equiv  \bm{\xi}_0 \cdot \bm{\xi}$. We choose the north pole of
$S_{n-1}$ to be $M_0$ and $\theta $ is thus the colatitude of point $M$ in
spherical coordinates.
We define
\begin{equation}
\langle \cos \theta \rangle (t) 
= \int_{S_{n-1}}
\rho(M,t) \; \cos \theta \; d\tau(M) \;
\end{equation}
where $d\tau(M)$ is the infinitesimal volume element in $S_{n-1}$. It follows
from  eq.\ (\ref{diff}) that
\begin{eqnarray}
\label{glo}
\frac{\partial}{\partial t }\langle \cos \theta \rangle &=&
\int_{S_{n-1}} D \cos \theta \; \Delta^{S_{n-1}}\rho(M,t) \; d\tau(M)
 \nonumber \\ 
&=&\int_{S_{n-1}} D \rho(M,t) \; \Delta^{S_{n-1}} \cos \theta \; d\tau(M) \; \;,
\end{eqnarray}
where we have  made use of Green's theorem in  $S_{n-1}$.\cite{Rosenberg}
 At this point we
recall that the Laplacian $\Delta^{E_{n}}$ in $E_n$ may be decomposed as the sum
\cite{Vilenkin}
\begin{eqnarray}
\label{Lap}
\Delta^{E_{n}}&=&\Delta^R + \Delta^{S_{n-1}} \nonumber \; ,\\
\Delta^R &=&\frac{1}{R^{n-1}} \frac{\partial}{\partial R}R^{n-1}
\frac{\partial}{\partial R} \; ,
\end{eqnarray}
where $\Delta^R$ acts only on the variable $R$ and $\Delta^{S_{n-1}}$,
which acts only on angular variables,  is called
the angular part of the Laplacian or, alternatively, the Laplace-Beltrami
operator.\cite{Vilenkin} 
From these remarks, the action of
$\Delta^{S_{n-1}}$ on $\cos\theta$ will be easily obtained. In  one hand 
we have obviously
$\Delta^{E_{n}}(\bm{OM}\cdot \bm{\xi}_0) =0$; however, on the other hand,
it follows from eqs.\ (\ref{Lap}) that 
$\Delta^{E_{n}}(\bm{OM}\cdot \bm{\xi}_0) \equiv \Delta^{E_{n}}(R\cos\theta)=
(n-1) \cos\theta/R  + R \Delta^{S_{n-1}}\cos\theta$ whence
 $\Delta^{S_{n-1}}\cos\theta=-(n-1)\cos\theta/R^2$ yielding,
 when inserted in eq.\ (\ref{glo}), to the simple equation
 \begin{equation}
 \frac{\partial}{\partial t }\langle \cos\theta \rangle 
 +(n-1) \frac{D}{R^2} \langle \cos\theta \rangle 
 =0 \; ,
 \end{equation} 
the solution of which being of course given by eq.\ (\ref{cos}).
 
\section{The general solution}
\label{general}
We shall make extensive use of the properties of the hyperspherical harmonics
to solve eq.\ (\ref{diff}). A short r\'esum\'e on these functions might prove
useful for the reader. We will adopt the definitions and notations of the 
chapter\ IX of the classical textbook by Vilenkin.\cite{Vilenkin} In the case
$n=4$ this yields to a definition of the harmonics which is slightly different
from that which we used  previously.\cite{Caillol2,Caillol3,Caillol4}

To avoid confusion we shall denote by $\mathcal{S}_{n-1}$ 
the hypersphere of radius $R=1$ (When, incidentally,
we talk about the volumes of the spaces
$S_{n-1}$ or $\mathcal{S}_{n-1}$ we will mean in fact the areas
 of these spheres considered as manifolds of $E_n$).
The spherical coordinates of the unit vector
$\bm{\xi} = \bm{OM}\equiv (\xi_1,\ldots,\xi_n)$ of 
$\mathcal{S}_{n-1}$  are defined by the relations
\begin{eqnarray}
\label{spheri}
\xi_1 &=&  \sin \theta_{n-1} \ldots  \sin \theta_{2} \sin \theta_{1}
 \nonumber  \; ,\\
\xi_2 &=&  \sin \theta_{n-1} \ldots  \sin \theta_{2} \cos \theta_{1} \; , \; \ldots  
\nonumber \; , \\
\xi_{n-1} &=&  \sin \theta_{n-1} \cos \theta_{n-2} \nonumber \; ,\\
\xi_{n} &=&  \cos \theta_{n-1} \; ,
\end{eqnarray}
where $0\leq \theta_{1} < 2\pi$ and  $0\leq \theta_{k} < \pi$ for $k\neq 1$.
Note that $\theta_{n-1}$ is the colatitude of point $M$ and was denoted 
by $\theta$ in section\ \ref{simple}.
The integration measure in $\mathcal{S}_{n-1}$ will be defined as
\begin{equation}
d\bm{\xi} \equiv \frac{1}{A_{n-1}} 
\sin^{n-2} \theta_{n-1} \ldots \sin \theta_{2} \; d\theta_{1} \ldots
d\theta_{n-1}
\; ,
\end{equation}
where $A_{n-1}=2\pi^{n/2}/\Gamma(n/2)$ is the surface of the sphere
$\mathcal{S}_{n-1}$.
With this normalization
\begin{equation}
\int_{\mathcal{S}_{n-1}} d\bm{\xi}=1 \; , 
\end{equation}
and the infinitesimal volume element in $S_{n-1}$ is given by 
$d\tau=R^{n-1}A_{n-1}d\bm{\xi}$.
The expression of the Laplacian 
$\Delta^{\mathcal{S}_{n-1}}\equiv R^2 \;\Delta^{S_{n-1}}$ 
in spherical coordinates reads as
\begin{eqnarray}
\label{lapspheri}
\Delta^{\mathcal{S}_{n-1}}&=&  \frac{1}{\sin^{n-2} \theta_{n-1}}
\frac{\partial}{\partial \theta_{n-1}} \sin^{n-2} \theta_{n-1}
\frac{\partial}{\partial \theta_{n-1}} + \nonumber \\
&+& \frac{1}{\sin^2 \theta_{n-1}  \sin^{n-3} \theta_{n-2}}
\frac{\partial}{\partial \theta_{n-2}} \sin^{n-3} \theta_{n-2}
\frac{\partial}{\partial \theta_{n-2}} + \ldots \nonumber \\
&+&\frac{1}{\sin^2 \theta_{n-1}\ldots \sin^2 \theta_2} 
\frac{\partial^2}{\partial \theta_1^2} \; .
\end{eqnarray}

The \mbox{(hyper)spherical} harmonics $\Xi_{L,\bm{K}}(\bm{\xi})$ 
are defined to be the eigenvectors of the operator 
$\Delta^{\mathcal{S}_{n-1}}$. The spectrum of $\Delta^{\mathcal{S}_{n-1}}$
is given by  the integers $\lambda(n,L)=-L(L+n-2)$ where
$L=0,1, \ldots$ is a positive integer. 
One can show that $\bm{K}\equiv (k_1, \ldots,\pm
k_{n-2})$ with $ L \geq k_1 \geq k_2 \geq k_{n-2} \geq 0$.
Therefore there are
exactly $h(n,L)=(2L+n-2)(n+L-3)!/(n-2)!L!  \;$ distinct harmonics 
$\Xi_{L,\bm{K}}(\bm{\xi})$ corresponding to the eigenvalue $\lambda(n,L)$.
The  $\Xi_{L,\bm{K}}(\bm{\xi})$ constitue a complete basis set for expanding
any square integrable function $f \in \mathcal{L}^2 [\mathcal{S}_{n-1}]$
defined on $\mathcal{S}_{n-1}$. An explicit expression of  
$\Xi_{L,\bm{K}}(\bm{\xi})$ will be of little use and can be found in the
Vilenkin.\cite{Vilenkin} As the usual spherical harmonics 
the $\Xi_{L,\bm{K}}(\bm{\xi})$ satisfy to the following
properties:\cite{Vilenkin}
\begin{itemize}
\item{(i)} orthogonality
\begin{equation}
\label{ortho}
\int_{\mathcal{S}_{n-1}} \Xi_{L,\bm{K}}^{\star}(\bm{\xi})\;
 \Xi_{L',\bm{K}'}(\bm{\xi}) \; d\bm{\xi} =\delta_{L,L'} \;
 \delta_{\bm{K},\bm{K}'}
 \; ,
\end{equation} 
\item{(ii)} completeness
\begin{equation}
\label{compl}
\sum_{L,\bm{K}} \Xi_{L,\bm{K}}^{\star}(\bm{\xi}) \;
 \Xi_{L,\bm{K}}(\bm{\xi}')=\delta^{\mathcal{S}_{n-1}}(\bm{\xi},\bm{\xi}') \; ,
\end{equation}
where
$ \delta^{\mathcal{S}_{n-1}}$ is the Dirac distribution for 
the unit hypersphere defined as
\begin{equation}
\int_{\mathcal{S}_{n-1}} f(\bm{\xi}) \; \delta^{\mathcal{S}_{n-1}}
(\bm{\xi},\bm{\xi}')
\; d\bm{\xi}= f(\bm{\xi}'). 
\end{equation}
\item{(iii)} addition theorem
\begin{equation}
\label{add}
\sum_{\bm{K}} \Xi_{L,\bm{K}}^{\star}(\bm{\xi})\;
 \Xi_{L,\bm{K}}(\bm{\xi}')=
 \frac{2L+n-2}{n-2} \;
 C_{L}^{n/2-1}
 (\bm{\xi}\cdot \bm{\xi}') \; ,
\end{equation}
where the dot in the RHS denotes the usual  scalar product in $E_n$, i.e 
$\bm{\xi}\cdot \bm{\xi}'=\cos\psi$ where $\psi$ is the angle between the two
unit vectors $\bm{\xi}$ and $\bm{\xi}'$, and $C_{L}^{n/2-1}$ is a Gegenbauer
polynomial.
The Gegenbauer polynomials $C_L^p$ are a generalization of Legendre
polynomials. $C_L^{p}$ is defined as the coefficient of $h^L$ in the
power-series expansion of the function 
\begin{equation}
\label{gene} 
(1-2th+h^2)^{-p}=\sum_{L=0}^{\infty} C_L^{p}(t)h^L \;.
\end{equation}
\end{itemize}
We have now in hand all the tools to solve eq.\ (\ref{diff}).
We first expand $\rho(M,T)$ in terms of the $\Xi_{L,\bm{K}}$,
\begin{equation}
\label{harm}
\rho(M,t)=\sum_{L=0}^{\infty} \sum_{\bm{K}} \rho_{L,\bm{K}}(t) \;
 \Xi_{L,\bm{K}}(\bm{\xi}) \; 
\end{equation}
and then insert the expansion\ (\ref{harm}) in eq.\ (\ref{diff}).
Making use of the orthogonal properties of the $\Xi_{L,\bm{K}}$ yields an
infinite system of non-coupled equations
\begin{equation}
\label{diff2}
\left(\frac{\partial}{\partial t} +D \; \frac{L(L+n-2)}{R^2} \right)
 \rho_{L,\bm{K}}(t)=0 \; ,
\end{equation}
the solution of which reads obviously as 
\begin{equation}
\label{solu}
\rho_{L,\bm{K}}(t)=
\rho_{L,\bm{K}}(0) \exp\left(-D L\left(L+n-2\right)t/R^2\right) \; .
\end{equation}
We shall denote by $\rho(M,t\vert M_0,0)$ the solution of\ (\ref{diff})
corresponding to
the initial condition $\rho(M,0)=\delta^{S_{n-1}}(M,M_0)\equiv 
\delta^{\mathcal{S}_{n-1}}(\bm{\xi}
,\bm{\xi}_0)/ R^{n-1} A_{n-1}$, i.e.
the solution of 
\begin{equation}
\label{Green}
\mathcal{D}\rho(M,t\vert M_0,0)=\delta(t) \; 
 \frac{1}{ R^{n-1} A_{n-1}}  \;\delta^{\mathcal{S}_{n-1}}(\bm{\xi}
,\bm{\xi}_0)
\; .
\end{equation}
It follows readily from eqs.\ (\ref{solu}) and\ (\ref{compl}) that the Green
function $\rho(M,t\vert M_0,0)$ can be expressed as
\begin{eqnarray} 
\label{Green-solu}
\rho(M,t\vert M_0,0)&=& 0 \; \; (t<0) \nonumber \\
\rho(M,t\vert M_0,0) &= & 
\frac{1}{R^{n-1} A_{n-1}} \sum_{L=0}^{\infty} \sum_{\bm{K}} \Xi_{L,\bm{K}}^{\star}(\bm{\xi}_0) \;
\Xi_{L,\bm{K}}(\bm{\xi}) \; \nonumber \\ 
& \times & \exp\left(-D L\left(L+n-2\right)t/R^2\right) 
\; \; (t>0)
\; .
\end{eqnarray}
Eq.\ (\ref{Green-solu}) can be further simplified with the help of the addition
theorem\ (\ref{add})  yielding our final result
\begin{eqnarray}
\label{Green-solu2}
\rho(M,t\vert M_0,0)&=&
\frac{1}{R^{n-1} A_{n-1}} \sum_{L=0}^{\infty} 
\frac{2L+n-2}{n-2}
C_{L}^{n/2-1}(\bm{\xi}_{0} \cdot \bm{\xi}) \nonumber \\
&\times&
\exp\left(-D L\left(L+n-2\right)t/R^2\right) \; \; (t>0) \; .
\end{eqnarray}

Some comments on eq.\ (\ref{Green-solu2})  are in order.

(i) The solution\ (\ref{Green-solu2}) is invariant under rotation about the axis
$\bm{\xi}_0$ as expected. Eq.\ (\ref{Green-solu2}) is a  generalization for all
$n$ of Debye's result which corresponds to the case
$n=3$.\cite{Debye,Berne} In this case the Gegenbauer polynomials 
reduce to the Legendre
polynomials $P_L$. In the case $n=4$ we recover the result
 of ref.~\onlinecite{Caillol1} where the Gegenbauer polynomials reduce to Tchebycheff
polynomials of second kind. To be more precise, recall that \cite{Vilenkin}
\begin{eqnarray}
C_{L}^{1/2} (\cos \theta ) &=& P_L (\cos \theta )  \; \; (n=3)  \\
C_{L}^{1} (\cos \theta ) &=& \frac{\sin (L+1)\theta}{\sin \theta} \; \; (n=4)
\end{eqnarray}

(ii) Note that $\int_{S_{n-1}} \rho(M,t\vert M_0,0) d\tau = 
\rho_{0,\bm{0}}=1$,  
i.e. the probability is conserved, and that we also have
$
\lim_{t \to + \infty} \rho(M,t\vert M_0,0)=1/\Omega_{n-1}$, 
where $\Omega_{n-1}=R^{n-1} A_{n-1}$ is the volume of the space $S_{n-1}$,
i.e. the solution of the diffusive process is uniform after infinite time.

(iii) Let us define the time correlation functions
\begin{eqnarray}
\mathcal{F}_{L}^{n/2-1}(t) & \equiv &  \langle
 C_{L}^{n/2-1}\left(\bm{\xi}_0 \cdot \bm{\xi}\right) \rangle 
 \; \; (L\geq 1) \nonumber \\
&=& \int_{\mathcal{S}_{n-1}} d\bm{\xi}_0 \; 
\int_{\mathcal{S}_{n-1}} d\bm{\xi}\; \; 
 C_{L}^{n/2-1}(\bm{\xi}_0 \cdot \bm{\xi})
\; \widehat{\rho}(M,t\vert M_0,0) \; ,
\end{eqnarray}
where we have introduced the reduced Green's function 
$\widehat{\rho} \equiv R^{n-1} A_{n-1}\rho$. 
As a consequence of the orthogonality properties\ (\ref{ortho}) we find that
\begin{equation}
\label{D}
\mathcal{F}_{L}^{n/2-1}(t) 
  =  \frac{(n-3+L)!}{L! (n-3)!} 
  \exp \left(-D L\left(L+n-2\right)t/R^2\right) \; 
\end{equation} 
Special cases are of interest. Firstly, since for $L=1$ we have
$C_{1}^{n/2-1}(u)=(n-2)u$,  equation\ (\ref{D}) indeed gives back
eq.\ (\ref{cos}). For $n=3$ and $4$ we recover the results for the $3D$ and
$4D$ rotors \cite{Debye,Berne,Caillol1}
\begin{eqnarray}
\label{parti}
\langle P_L\left(\cos \theta \right) \rangle  & = & 
 \exp \left(-D L\left(L+1 \right)t/R^2\right) \; \; (n=3)  \\
\label{n4}
\langle \frac{\sin \left(L+1\right) \theta }{\sin \theta}\rangle 
 & = & (L+1)
 \exp \left(-D L\left(L+2\right)t/R^2\right) \; \; (n=4) 
 \; ,
\end{eqnarray}
Defining now the reorientational time $\tau_L^{n/2-1}$  as 
\begin{equation}
\label{tauL}
\tau_L^{n/2-1} = \int_0^{\infty} \frac{\mathcal{F}_{L}^{n/2-1}(t)} 
{\mathcal{F}_{L}^{n/2-1}(0)} \; dt  
       =  \frac{R^2}{D L (L+n-2)} \; ,
\end{equation}    
we have now at our disposal a Kubo formula for the diffusion coefficient
$D$.  Note the aesthetic relation
 $\tau_L^{n/2-1} /\tau_{L'}^{n/2-1}=L'(L'+n-2)/L(L+n-2)$ which
generalizes Debye's result  to arbitrary
dimensions.\cite{Debye,Berne} Since simulations of real $3D$
liquids or plasmas are feasible (moreover efficient) 
in $S_3$, \cite{Caillol2,Caillol3,Caillol4}
eq.\ (\ref{tauL})  is of prime
importance since it should allow the computation of the
self-diffusion coefficient of
such systems in the course of equilibrium molecular dynamics simulations.

(iv) Simulations of random walks on fractals inscribed in $S_{n-1}$
were reported recently.\cite{Jullien} While it is known that for random walks
on fractal clusters in Euclidean spaces the anomalous diffusion law
$\langle r^2(t)\rangle \varpropto t^{\beta}$ holds\cite{Alexander,Gefen} (where
$\beta$ is some exponent depending of the fractal dimensions of the cluster)
the numerical results of
ref.~\onlinecite{Jullien} give
evidence of the  law 
$\langle \cos \theta \rangle \propto \exp(-(t/\tau)^{\beta}) $,
i.e. a stretched
exponential relaxation, on the hypersphere. As suggested by Jund {\it et
al.}\cite{Jullien} the anomalous diffusion law for a RW on a fractal can
be understood by replacing the time $t$ by a fractal time $t^{\beta}$ in the
diffusion equation\ (\ref{diff}) and thus in the solution\ (\ref{Green-solu2}).
With this assumption we now have the ansatz
\begin{equation}
\left(\tau_L^{n/2-1} /\tau_{L'}^{n/2-1}\right)^{1/\beta}=L'(L'+n-2)/L(L+n-2) \; ,
\end{equation} 
which could be checked in numerical simulations.
\section{Random Walks in $S_3$}
\label{S3}
We specialize to the case $n=3$ and want to show that 
the expression\ (\ref{Green-solu2}) of 
$\rho(M,t\vert M_0,0)$ is equivalent to that of ref.\ [\onlinecite{Nissfolk}].
This can be done as follows. Let us rewrite the reduced
 $\widehat{\rho}= 2 \pi^2 R^3 \rho(M,t\vert M_0,0)$
as
\begin{equation}
\widehat{\rho}=\sum_{L=0}^{\infty} (L+1) \frac{  \sin(L+1) \psi}{\sin \psi}
\exp(-K \; L (L+2)) \; ,
\end{equation}
where $K=Dt/R^2$. {\em A priori} the angle $\psi$ is in the range $(0,\pi)$,
however since the function is formally even in $\psi$ we define 
$\widehat{\rho}(-\psi)=\widehat{\rho}(\psi)$
for negative angles. This gives us a periodic function of period $2 \pi$ defined
for all $\psi \in \R$. We introduce now the periodic function
\begin{equation}
F(\psi)=\int_{0}^{\psi} d \psi ' \;  \widehat{\rho}(\psi ') \sin(\psi ') 
\end{equation}
which can be rewritten after some algebra as
\begin{equation}
F(\psi)= F_0 -\frac{\exp{K}}{2} \sum_{p= -\infty}^{+\infty}
\exp\left(-K p^2\right) \exp (-i p \psi) 
\end{equation}
where $F_0$ is some unessential constant independent of angle $\psi$.
At this point we  recall Poisson
summation theorem which states that for any function $\varphi(x)$ holomorphic in
the strip $-a< \Im z <a$ one has
\begin{equation}
\label{Poisson}
\sum_{n= -\infty}^{+\infty} \varphi(x+2 n \pi)=
\frac{1}{2\pi} \sum_{p= -\infty}^{+\infty} e^{-ipx} \; 
\int_{-\infty}^{+\infty}  \varphi(y) e^{i p y} \;  dy \; . 
\end{equation}
Applying Poisson theorem for the Gaussian we get 
\begin{equation}
F(\psi)= F_0 - \frac{\sqrt{\pi}\exp{K}}{2 \sqrt{K}}
\sum_{n=-\infty}^{+\infty} \exp \left(  
-\frac{(\psi + 2n \pi)^2}{4K}\right) \; ,
\end{equation}
which, after differentiation yields for $\widehat{\rho}$ 

\begin{equation}
\label{niss}
\widehat{\rho}(\psi,t)= \frac{\sqrt{\pi}\exp{K}}{4K^{3/2} \sin \psi}
\sum_{n=-\infty}^{+\infty} (\psi + 2n \pi)
\exp \left(  
-\frac{(\psi + 2n \pi)^2}{4K}\right)  \; ,
\end{equation}
which coincides with the result of ref.~\onlinecite{Nissfolk}, apart  the
prefactor which is not specified.
Relations  similar to eq.\ (\ref{niss}) can be
obtained for \textit{even} values of $n$ (other than 4)
but we failed to get anything similar for odd $n$'s.

\begin{acknowledgments}
The author likes to thank I. Campbell for drawing his attention to
ref.~\onlinecite{Jullien} and for enlighting e-mail correspondence.
\end{acknowledgments}

\end{document}